\documentclass[twocolumn,showpacs,preprintnumbers,amsmath,amssymb]{revtex4}
\usepackage{graphicx}

\def\be{\begin{equation}}
\def\ee{\end{equation}}
\def\ba{\begin{eqnarray}}
\def\ea{\end{eqnarray}}
\def\go{\mathrel{\raise.3ex\hbox{$>$}\mkern-14mu
             \lower0.6ex\hbox{$\sim$}}}
\def\lo{\mathrel{\raise.3ex\hbox{$<$}\mkern-14mu
             \lower0.6ex\hbox{$\sim$}}}

\def\bx{{\bf x}}
\def\brph{{\bf r}_{\rm ph}}
\def\bn{{\bf n}}
\def\bb{{\bf b}}
\def\bm{{\bf m}}

\def\bv{{\bf v}}
\def\bk{{\bf k}}
\def\bs{{\bf s}}
\def\br{{\bf r}}
\def\bR{{\bf R}}
\def\bV{{\bf V}}
\def\bS{{\bf S}}
\def\bx{{\bf x}}

\begin{document}

\title{Effects of Gravitational Lensing and Companion Motion 
on the Binary Pulsar Timing}
\author{Roman R. Rafikov}
\email{rrr@cita.utoronto.ca}
\affiliation{CITA, McLennan Physics Labs, 60 St. George St., University of Toronto, Toronto, ON M5S 3H8, Canada}
\author{Dong Lai}
\email{dong@astro.cornell.edu}
\affiliation{Department of Astronomy, Cornell University, 
Ithaca, NY 14853}

\date{\today}

\begin{abstract}
The measurement of the Shapiro time delay in binary pulsar systems 
with highly inclined orbit can be affected both by the motion 
of the pulsar's companion because of the finite time it 
takes a photon to cross the binary, and by the gravitational light bending
if the orbit is sufficiently edge-on relative to the line of sight.
Here we calculate the effect of retardation due to the companion's
motion on various time delays in pulsar binaries, including the
Shaipro delay, the geometric lensing delay, 
and the lens-induced delays associated with the pulsar rotation.
Our results can be applied to systems so highly  
inclined that near conjunction gravitational lensing 
of the pulsar radiation by the companion becomes important 
(the recently discovered double pulsar system J0737-3039 may 
exemplify such a system). 
To the leading order, the effect of retardation is to shift all 
the delay curves backward in time around the orbit conjunction, without 
affecting the shape and amplitude of the curves. 
The time shift is of order the photon orbit crossing time,
and ranges from a second to a few minutes for the observed 
binary pulsar systems.
In the double pulsar system J0737-3039, the motion of the companion 
may also affect the interpretation of the recent correlated interstellar 
scintillation measurements.
Finally, we show that lensing sets an upper limit on the magnitude of 
the frame-dragging time delay caused by the companion's spin, and makes 
this delay unobservable in stellar-mass binary pulsar systems.
\end{abstract}

\pacs{95.30.Sf -- 95.85.Bh--97.60.Gb--97.60.Jd--97.60.Lf--97.80.-d}


\maketitle


\section{Introduction
\label{sect:intro}}


Timing study of binary radio pulsars has provided some of the best
tests of the general relativity (GR) to date.  Among the effects that
have been probed in different systems are gravitational redshift,
precession of the periastron and orbital decay due to gravitational
wave emission. Strong evidence for the geodetic precession -- GR
manifestation of the spin-orbital coupling of the binary -- has been
found in several systems (e.g., \cite{stairs}, \cite{hotan}).  In
binaries with highly inclined orbits, it has been possible to measure
the Shapiro delay caused by the propagation of the pulsar signal in
the companion's gravitational field (e.g., \cite{lyne},
\cite{splaver}, \cite{jacoby}).

In systems that have nearly edge-on orbital orientation, like 
the recently discovered double pulsar system J0737-3039 \cite{lyne},
one may also look for the effects related to the gravitational bending 
of the pulsar radio beam as it passes close to the companion around 
the moment of the pulsar's superior conjunction 
(\cite{schneider}, \cite{dorkop}, \cite{LR05}, \cite{RL05}). 
Light bending modifies the conventional Shapiro delay formula
widely used in pulsar timing studies and introduces additional 
geometric time delay (\cite{schneider}, \cite{LR05}). 
Bending also gives rise to a {\it lens-rotational delay} associated with 
the rotating beam of the pulsar and induces distortion of the 
pulse profile shape near conjunction \cite{RL05}. 

Our previous studies of the lensing effects in binary pulsar systems
(\cite{LR05}, \cite{RL05}) adopted a ``static lens'' approximation, in which 
the companion motion is neglected. 
Kopeikin \& Schafer \cite{kopshaf} showed that the orbital 
motion of the pulsar companion gives rise to a $v_c/c$ correction 
($v_c$ is the companion velocity) to 
the conventional Shapiro delay formula and such correction might be
observable under some circumstances. 
The work of \cite{kopshaf}, however, neglected light bending
and thus cannot be applied to the situation
where the light ray passes the companion within a few Einstein 
radii. On the other hand, only for such highly inclined systems 
are the lensing effects potentially observable (\cite{LR05}, \cite{RL05}).

In this paper we study the effect of companion motion on the 
time delays associated with the gravitational light bending 
near the companion. These include both the
lensing corrections to the delays independent of the pulsar 
spin -- Shapiro and geometric delays (see \S 
\ref{sect:timing_no_spin}), and the lens-rotational time 
delays (\S \ref{sect:spin}). Our formulae are general and valid
in both the ``strong lensing'' and the ``weak lensing''
regimes. We also provide a simple and general interpretation of the 
influence of the lens motion on the behavior of various timing signals in
binary pulsar systems (\S \ref{sect:implications}) 
which has not been previously discussed. 

Qualitatively, we expect 
that near binary conjunction the effect of moving lens
amounts to a time shift in various time delay curves.
It takes finite amount of time
for the pulsar signal to travel across the binary system during which
the position of its companion changes. Based on simple geometric
considerations, this time shift is $(M_p/M)a_\parallel/c$,
where $M_p$ is the pulsar mass, $M$ is the
total mass of the system, and $a_\parallel$ is a separation between
binary components projected along the line of sight 
[see eq. (\ref{eq:a_par})]. In the limit $M_p\to 0$
this retardation delay goes to zero since companion does not move.
Our explict calculations and general expressions of various
time delays confirm this simple expectation, see \S \ref{sect:implications}.

For completeness,  in \S \ref{sect:FD} we discuss the effect of frame dragging 
due to the companion's spin
on the time delay. This type of delay was previously investigated by 
\cite{laguna}, \cite{wex}, \cite{tart}, \cite{rugg}, who 
disregarded light bending in their studies. 
We incorporate the lensing effect in our calculation of the frame-dragging 
delay, and show (\S \ref{subsect:fd_interp}) that this delay is 
unobservable in the binary pulsar systems with 
stellar mass companions.

\section{Lensing by Moving Pulsar Companion
\label{sect:lensing}}

Consider a binary system consisting of a pulsar with mass 
$M_p$ and a companion with mass $M_c$
(the total mass is $M=M_p+M_c$).
Let $\br_p$, $\br_c$ be their position vectors 
with respect to the binary center-of-mass, and  
$\bR_p$, $\bR_c$ be their projections on the plane of the sky. 
We also define $\br=\br_p-\br_c$ (the position vector of the pulsar 
relative to its companion), $r_\parallel\equiv -\bn_0\cdot\br$
(the projection of $\br$ along the line of sight; here $\bn_0$ is 
the unit vector from the binary barycenter to the observer), and 
$\bR\equiv \bR_p-\bR_c=\bn_0\times 
(\br\times \bn_0)$ (the projection of $\br$ in the sky plane).
For a binary orbit characterized by the semimajor axis $a$, 
eccentricity $e$, longitude of periastron $\omega$ and inclination angle
$i$, we have
\ba
&&r=|\br|=a(1-e^2)/(1+e\cos\phi),\\
&&r_\parallel=r\sin i\sin \psi,\\ 
&&R=|\bR|=r(1-\sin^2 i\,\sin^2\psi)^{1/2},
\ea
where $\psi$ is the true anomaly measured from the ascending node of 
the pulsar and $\phi=\psi-\omega$ is the orbital true anomaly measured from 
the periastron. To describe gravitational lensing of the pulsar signal 
by its companion, it is also useful to introduce 
\ba
a_\parallel=a|\sin i|(1-e^2)/(1+e\sin\omega),
\label{eq:a_par}
\ea
which is the value of $r_\parallel$ at the superior conjunction 
of the pulsar ($\psi=\pi/2$).

A photon (radio pulse) emitted by the pulsar at time $t=t_e$ from the
position $\bR(t_e)$ passes through the lens plane at time 
$t_l\simeq t_e+r_\parallel/c$. In the absence of light bending, 
the impact parameter of the ray is 
\ba
\bb_0 &=&\bR_p(t_e)-\bR_c(t_l)=\bR(t_e)+\bR_c(t_e)-\bR_c(t_l)\nonumber\\
&\simeq & \bR(t_e)-r_\parallel \frac{\bV_c(t_e)}{c}
= \bn_0\times\left[\left(\br-r_\parallel{\bv_c\over c}
\right)\times\bn_0\right].
\label{eq:bb_0}
\ea
where $\bV_c\equiv d\bR_c/dt=\bn_0
\times(\bv_c\times \bn_0)$ is the projection of the companion velocity
$\bv_c$ in the plane of the sky. 
In terms of the binary orbital parameters we have
\ba
&& b_0^2=\left[r\cos\psi-d(\sin\psi+e\sin\omega)\right]^2
\nonumber\\
&& +\left[r\sin\psi\cos i+d\cos i(\cos\psi+e\cos\omega)\right]^2.
\label{eq:b0^2}
\ea
Here, to characterize the effect of lens motion, 
we have introduced the parameter $d$ given by
\ba
d\equiv a_\parallel\frac{\Omega_b a_c}{c\sqrt{1-e^2}}=
a_c\frac{\Omega_b a\sin i}{c(1+e\sin\omega)},
\label{eq:d}
\ea
where $a_c=a(M_p/M)$ is the companion's semi-major axis
and $\Omega_b=(GM/a^3)^{1/2}$ is the binary angular 
frequency. Note that in eq.~(\ref{eq:b0^2}) and the remainder of the
paper, all variables are evaluated at the time of pulse emission
$t=t_e$.

Because of light bending, the actual impact parameter of the ray
(i.e. the minimum-approach distance between the radio beam
and the companion), $b$, differs from $b_0$. The bending angle, including 
the leading-order correction due to the companion motion, is 
(e.g., \cite{pyne}, \cite{wuck})
\ba
\alpha=\left(1-\frac{\bn_0\cdot \bv_c}{c}\right)
\frac{4GM_c}{c^2b}, 
\label{eq:angle}
\ea
The $v_c/c$ correction to the classical light-bending
formula only slightly affects the value of the companion's
Einstein radius, and will be dropped in the remainder of the
paper. We are only interested in lensing around the superior
conjunction of the pulsar, during which $b_0\sim R\ll r_\parallel
\simeq a_\parallel$ is well satisfied. 
Lensing gives rise to two images (positive and negative) of the source
located at $\bb_\pm=b_\pm\bb_0/b_0$ in the plane of the sky 
with respect to the companion's position. From the lensing equation,
$b-b_0=\alpha a_\parallel$, we find the impact parameters of the two
images:
\be
b_\pm=\frac{1}{2}
\left(b_0\pm\sqrt{b_0^2+4R_E^2}\right),
\label{eq:b_def}
\ee
where $R_E$ is the Einstein radius given by
\be
R_E=(2R_g a_\parallel)^{1/2},~~~R_g=2GM_c/c^2.
\label{eq:ein_rad}
\ee
The image amplification $A_\pm=d(b_\pm^2)/d(b_0^2)$ is given by
the standard expression 
\be
A_\pm={u^2+2\over 2u\sqrt{u^2+4}}\pm\frac12,\qquad {\rm with}~~
u=b_0/R_E.
\label{eq:mag}
\ee

Clearly, the motion of the companion affects the lensing images
only through the ``retardation'' term in eq.~(\ref{eq:bb_0}),
arising from the finite time $r_\parallel/c$ that photons spend
traveling between the pulsar and the lens plane of the companion.


\section{Geometric and Shapiro Delays
\label{sect:timing_no_spin}}


Lai \& Rafikov \cite{LR05} gave the expressions for the geometric delay and the 
lensing-corrected Shapiro delay for highly inclined binary pulsar systems
based on the static approximation. Here we revise these 
expressions taking into account the motion of the companion.

To order ${\cal O}(v_c/c)$, the expression for the geometric delay 
assumes the usual form (see Appendix A)
\ba
(\Delta t)_{\rm geom}=\frac{R_g}{c}
\left(\frac{\Delta b_\pm}{R_E}\right)^2,
\label{eq:geom}
\ea
where the image displacement is given by 
\be
\Delta b_\pm=b_\pm-b_0 =
\frac{1}{2}
\left(\pm\sqrt{b_0^2+4R_E^2}-b_0\right).
\label{eq:eps}
\ee
The leading-order $v_c/c$ correction to $(\Delta t)_{\rm geom}$ 
enters through the expression for $b_0$ [see eq.~(\ref{eq:b0^2})].
Compared to the the static limit \cite{LR05} the
only difference is in using $b_0$ instead of $R$ in eq.~(\ref{eq:eps}).

The easiest way to illustrate how the lens motion affects 
the Shapiro delay is to switch to a reference frame co-moving with 
the companion\footnote{Although $\bv_c$ is not constant, 
as long as the orbital period of the system is longer than the 
photon propagation time across the binary, the companion's 
acceleration can be neglected, see \cite{kopshaf} .}.
In this frame, the companion's gravitational field is static and 
the expression for the Shapiro delay derived 
in \cite{LR05} [see their eq. (5)] can be used. The expression involves
$r_\parallel$ and $b$. Because of the aberration of light, the 
direction of photon propagation in the comoving frame ${\bf n}^\prime$ 
is given by $\bn^\prime\simeq \bn_0-\bV_c/c$. Thus,
instead of $r_\parallel$ one has to use 
$r_\parallel^\prime=-\bn^\prime\cdot\br= r_\parallel+\bR\cdot\bV_c/c$.
Also, the difference between $b_\pm'$ (the impact parameter of the ray 
in the comoving frame) and $b_\pm$ is proportional to $(v_c/c)^2$.
Thus, including the leading order $v_c/c$ correction, the gravitational
(Shapiro) delay is given by (cf. eq.~(5) of \cite{LR05})\footnote{Note that 
we include the constant term $(R_g/c)\ln\left[a(1-e^2)\right]$ 
in eq.~(\ref{eq:Shapiro}) so that in the static, no-lensing limit
it reduces to the conventional expression (e.g., \cite{bland})
$$(\Delta t)_{\rm grav}=(R_g/c)\ln\left[{1+e\cos\phi
\over 1-\sin i\sin(\phi+\omega)}
\right].$$}
\ba
&&(\Delta t)_{\rm grav}=-\left(1-{\bn_0\cdot\bv_c\over c}\right)\frac{R_g}{c}
\nonumber\\
&& \times\ln\left[\sqrt{(r_\parallel+\bR\cdot \bV_c/c)^2+b_\pm^2}-
r_\parallel-\bR\cdot \bV_c/c\right]\nonumber\\
&&+{R_g\over c}\ln\left[a(1-e^2)\right].
\label{eq:Shapiro} 
\ea
Note that the prefactor $(1-\bn_0\cdot\bv_c/c)$,
which was absent in the original formula of \cite{LR05}, is 
not captured in our simple analysis given above.
A formal derivation of eq.~(\ref{eq:Shapiro}) is given in 
Appendix A.

In the ``weak lensing'' limit, when $b_0\gg R_E$ [or 
equivalently, $(\Delta\psi)^2+(\Delta i)^2\gg 
(R_E/a_\parallel)^2\sim (v/c)^2$, where
$\Delta \psi=\psi-\pi/2$ and $\Delta i=i-\pi/2$], we have
for the positive image\footnote{The negative image is highly demagnified
and is unobservable.},
$\Delta b_+\simeq R_E^2/b_0=2a_\parallel (v/c)(R_E/b_0)\ll
a_\parallel (v/c)$, where $v\simeq (GM/a_\parallel)^{1/2}$. Thus, 
$\bb_+\simeq \bR-r_\parallel\bV_c/c$.
Keeping only terms up to order ${\cal O}(v_c/c)$, the numerator of 
the expression inside the logarithm in eq.~(\ref{eq:Shapiro})
reduces to $r-r_\parallel-{\bR\cdot \bV_c}/{c}=
r+\br\cdot\left[\bn_0-\bn_0\times(\bv_c\times\bn_0)/c\right]$.
As a result, in the absence of lensing, eq.~(\ref{eq:Shapiro}) becomes
\ba
&& (\Delta t)_{\rm grav}=-\left(1-{\bn_0\cdot\bv_c
\over c}\right)\,\frac{R_g}{c}\nonumber\\
&& \times\ln
\left[r+\bn_0\cdot\br-{\bn_0\times\bv_c\over c}\cdot(\bn_0\times\br)\right]
\nonumber\\
&& +{R_g\over c}\ln\left[a(1-e^2)\right].
%
\label{eq:Wex}
\ea
which agrees with the no lensing result obtained in \cite{kopshaf} (see
also \cite{will}).

Not too far from the binary conjunction, when $R\simeq a_\parallel
[(\Delta\psi)^2+(\Delta i)^2]^{1/2}\ll a_\parallel$,
eq.~(\ref{eq:Shapiro}) 
reduces to 
\ba
&& (\Delta t)_{\rm grav}=-\left(1-{\bn_0\cdot\bv_c\over c}\right)
\frac{R_g}{c}\ln
\left(\frac{b_\pm^2 f}{2r_\parallel}\right)\nonumber\\
&& +{R_g\over c}\ln\left[a(1-e^2)\right].
\label{eq:Shap_conj}
\ea
where $f\simeq 1-\bR\cdot\bV_c/(cr_\parallel)\simeq 1$.
Note that the regions of validity of the limiting cases
described by eqs.~(\ref{eq:Shap_conj}) and (\ref{eq:Wex}) have a 
significant interval of overlap 
[$(v/c)^2\ll(\Delta\psi)^2+(\Delta i)^2\ll 1$].


\section{Lens-Rotational Delays
\label{sect:spin}}


Because of light bending of the pulsar signal by the companion,
the direction of emission ${\bf n}$ at the pulsar position differs from 
${\bf n}_0$, with the deflection vector
$(\Delta\bn)_L=\bn-\bn_0$ given by 
\be
(\Delta\bn)_L={\Delta\bb_\pm\over a_\parallel}
=\left({\Delta b_\pm\over b_0}\right)
{\bn_0\times [(\br-a_\parallel\bv_c/c)\times \bn_0]
\over a_\parallel},
\label{eq:deln_L}
\ee
where $\Delta\bb_\pm=\bb_\pm-\bb_0=\Delta b_\pm (\bb_0/b_0)$
[see eqs.~(\ref{eq:bb_0}) and (\ref{eq:eps})].
Again, we are interested in the lensing effect only 
around the superior conjunction of the pulsar, so that 
$r_\parallel\simeq a_\parallel$. Since pulsar signals are due to the 
beamed emission of the neutron star (as opposed to radial pulsation),
such a change in the emission direction gives rise to 
time delays and pulse profile distortion which depend on
the pulsar spin period (\cite{RL05})\footnote{The subscript ``L'' in 
$(\Delta\bn)_L$ serves as a reminder that the effect is due to
lensing. The orbital motion of the pulsar also gives rise to
aberration of the emission direction in the pulsar's rest frame;
such ``aberration delays'' (\cite{smarr}, \cite{DD}, 
\cite{RL05}) are not affected directly by the motion of the companion,
and are not discussed in this paper.}.

In the following we will use the description of the pulsar spin 
geometry from \cite{DT} (see their Fig.~1) in which 
the pulsar spin axis $\bs_p$ (the unit vector along the pulsar 
spin direction) is specified by two angles: $\zeta_p$ is 
the angle between $\bs_p$ and the light-of-sight vector $\bn_0$,
and $\eta_p$ is the angle between the ascending node of the orbit and the 
projection of $\bs_p$ on the sky plane. 


\subsection{Longitudinal Time Delay
\label{subsect:longitudinal}}

Variation of the emission direction $(\Delta \bn)_L$ causes a shift in the 
equatorial {\it longitude} $\Phi$ of the emission direction in the 
corotating frame of pulsar (counted in the direction of rotation):
\be
\Delta\Phi=\frac{(\Delta \bn)_L \cdot ({\bf s}_p\times{\bf n}_0)}
{|{\bf s}_p\times{\bf n}_0|^2}.
\label{eq:delta_phi}
\ee
This longitudinal shift corresponds to a change of the emission phase, 
leading to a {\it longitudinal} time delay $(\Delta t)_L=\Delta\Phi/\Omega_p$
($\Omega_p$ is the angular frequency
of the pulsar; the time delay is positive for signal arriving later).
This delay is independent of the specific emission pattern (as 
long as it is rigidly rotating around ${\bf s}_p$) and is given by 
(cf. \cite{RL05})
\begin{eqnarray}
(\Delta t)_L=
\left({\Delta b_\pm\over b_0}\right){ (\br-a_\parallel\bv_c/c)
\cdot (\bs_p\times\bn_0)\over
\Omega_p a_\parallel |\bs_p\times\bn_0|^2}
\label{eq:lensing_delay}
\end{eqnarray}
The longitudinal time delay shifts the pulse profile uniformly in time 
without affecting its shape. Evaluating eq.~(\ref{eq:lensing_delay})
explicitly using the pulsar spin geometry described above,
we find
\begin{widetext}
\ba
&& (\Delta t)_L = -\left({\Delta b_\pm\over b_0}\right)
\nonumber\\
&& \times\left[
\left({r\over a_\parallel}\right)
{\sin\eta_p\,\cos\psi-\cos i\,\cos\eta_p\,\sin\psi \over \Omega_p 
\sin\zeta_p}-\left({d\over a_\parallel}\right)
{\sin\eta_p(\sin\psi+e\sin\omega)+\cos i\cos\eta_p(\cos\psi+e\cos\omega) 
\over \Omega_p \sin\zeta_p}\right],
\label{eq:dtl}
\ea
\end{widetext}
where $b_0$ is given by eq.~(\ref{eq:b0^2}). Setting $d=0$ recovers
the expression for the longitudinal lens-rotational delay obtained in
\cite{RL05} under static approximation.


\subsection{Pulse Profile Variation and Latitudinal Time Delay
\label{subsect:latitudinal}}

Variation of the pulse emission direction $(\Delta \bn)_L$ also results
in the change of {\it colatitude} $\zeta$ of the emission vector:
\ba
&& (\Delta\zeta)_L=-\frac{{\bf s}_p\cdot (\Delta \bn)_L}
{|{\bf s}_p\times{\bf n}_0|}\nonumber\\
&& = \left({\Delta b_\pm \over b_0}\right)
{\bn_0\times (\br-a_\parallel \bv_c/c)
\over a_\parallel}
\cdot{(\bs_p\times\bn_0)\over |\bs_p
\times\bn_0|}.
\label{eq:delta_lambda_L}
\ea 
In general, any shift of the (co)latitude of the emission vector leads 
to a pulse profile variation because the pulsar radio emission 
is thought to be tied to a specific geometric pattern on a 
celestial sphere rigidly rotating around $\bs_p$. Since the 
efficient particle acceleration leading to radio 
emission is thought to take place at the pulsar magnetic polar caps, 
the radio emission pattern is expected to be connected with the 
position of the pulsar magnetic axis ${\bf m}$.  

Here, for illustrative purposes, we assume that the radio emission pattern
forms a set of coaxial cones with circular cross sections and 
symmetry axes coincident with the magnetic axis ${\bf m}$, which
is inclined at an angle $\alpha$ with respect to the pulsar spin axis 
$\bs_p$ (see Fig.~3 of \cite{RL05} for details of the adopted pulsar 
spin-magnetic geometry). This emission pattern forms the basis of 
the rotating vector model (RVM, see \cite{radha}) commonly adopted 
for inferring the pulsar spin-magnetic orientation from the 
pulse profile and polarization data. In the course
of pulsar rotation around $\bs_p$, a given emission cone is
traversed by the line of sight to the observer $\bn_0$  
only if the cone's half-opening angle 
$\rho$ satisfies $\rho>|\zeta_p-\alpha|$.
Crossing the edges of the cone by $\bn_0$ leads to the 
leading and trailing emission episodes observed at Earth.

In the RVM, the latitudinal shift $(\Delta\zeta)_L$ [due to
$\bn_0\rightarrow\bn_0+(\Delta\bn)_L$] results in a variation of the full width
$2\Phi_0$ of the pulse profile (corresponding to a given 
emission cone), with \cite{RL05}
\be
\Delta\Phi_0=-{(\Delta\zeta)_L\over \sin\zeta_p\tan\chi_0}.
\label{eq:delphi0}
\ee
Here $\chi_0$ is the angle (on the celestial sphere) between the 
arc connecting $\bn_0$ and $\bs_p$ and the arc connecting $\bn_0$ 
and $\bm$ at the edges of the pulse (see Figs. 3 \& 4 of \cite{RL05}). 
In the framework of the RVM, $\chi_0$ is directly related 
to the position angle of the linear polarization of the pulsar 
emission, and is given by \cite{komes}
\be
\tan\chi_0=\frac{\sin\alpha\sin\Phi_0}{\cos\alpha\sin\zeta_p-
\cos\Phi_0\sin\alpha\cos\zeta_p}.
\label{eq:chi}
\ee

The pulse contraction/expansion $\Delta\Phi_0$ directly
translates to a {\it latitudinal delay} of the arrival time of various pulse
{\it components}. This delay is $(\Delta t)_L^{\rm (lat)}=\Delta\Phi_0/\Omega_p$
for the leading component of the pulse profile\footnote{
For the trailing component, the latitudinal delay is 
$-\Delta\Phi_0/\Omega_p$. Hereafter our formulae refer to the 
leading component.}.
Combining eqs.~(\ref{eq:delta_lambda_L}) and (\ref{eq:delphi0}) 
we find
\begin{widetext}
\ba
(\Delta t)_L^{\rm (lat)}=-\frac{(\Delta \zeta)_L}
{\Omega_p\sin\zeta_p\tan\chi_0}=
-\left({\Delta b_\pm\over b_0}\right)
\nonumber\\
\times \left[
\left({r\over a_\parallel}\right)
{\cos\eta_p\,\cos\psi+\cos i\,\sin\eta_p\,\sin\psi \over 
\Omega_p\sin\zeta_p\tan\chi_0}+\left({d\over a_\parallel}\right)
{\cos i\sin\eta_p(\cos\psi+e\cos\omega)-\cos\eta_p(\sin\psi+e\sin\omega) 
\over \Omega_p\sin\zeta_p\tan\chi_0}\right].
\label{eq:delta_t_L^lat}
\ea
\end{widetext}
Setting $d=0$ in eq.~(\ref{eq:delta_t_L^lat}) reproduces equation (23) 
of \cite{RL05}, which was derived in the static approximation.
 
Since $(\Delta t)_L^{\rm lat}$ depends through $\chi_0$ on phases 
$\Phi_0$ of different pulse components, the latitudinal lensing 
time delay leads 
to inhomogeneous dilation/contraction of the pulse profile
around the superior conjunction of the pulsar\footnote{Although 
our formulae are based on the RVM,
similar latitudinal pulse shape variation and delays
should be exhibited by all non-pathological emission patterns,
not just that assumed by the RVM.}.
This is in contrast to the longitudinal shift of the emission vector,
which only causes homogeneous displacement of the whole pulse profile in
time without introducing any distortions of its shape. 
This difference of behaviors led \cite{RL05} to propose timing of the 
{\it individual features} of the pulse profile as a means of detecting 
the latitudinal shift. 
Such a procedure yields a larger information content than the 
standard timing of the {\it whole} pulse profile (assumed to have 
fixed shape) which can only reveal the longitudinal time delay.

\section{Frame-Dragging Time Delay
\label{sect:FD}}

Frame-dragging delay $(\Delta t)_{\rm FD}$ owes its existence to
the gravitomagnetic effect of the companion spin angular momentum $\bS_c$. 
In order of magnitude, the metric perturbation due to $\bS_c$
is $\sim S_c/r^2$ (in units where $G=c=1$), the additional time delay
for a ray with impact parameter $b$ is then $(\Delta t)_{\rm FD}\sim
S_c/b$. Given the small magnitude of this delay,
in its subsequent consideration we will neglect the effect of
the companion motion and will use static approximation accounting 
only for lensing effects. We have (see e.g., \cite{wex})
\ba
(\Delta t)_{\rm FD}=-\frac{2G}{c^4}\int d\br_{ph}\cdot
\frac{\bS_c\times(\br_{ph}-\br_c)}{|\br_{ph}-\br_c|^3},
\label{eq:delay_formula}
\ea
where $\br_{ph}$ is the position vector for the photon 
trajectory (see also Appendix \ref{ap:Shapiro}).
Assuming that the observer is at infinite distance from the system and
evaluating the integral in (\ref{eq:delay_formula}) we find
\ba
(\Delta t)_{\rm FD}=-\frac{2G}{c^4}\frac{\bn_0\cdot(\bS_c\times\bR)}
{b_\pm R}\left[1+\frac{r_\parallel}
{(r_\parallel^2+b_\pm^2)^{1/2}}\right].
\label{eq:delay}
\ea
As $(\Delta t)_{\rm FD}$ may only be 
noticeable in nearly edge-on systems and at the conjunction,
one can use $b_\pm\ll r_\parallel$. 
We specify the orientation of the companion's spin vector $\bS_c$ 
by the angles $\zeta_c$ and $\eta_c$ (analogous to angles $\zeta_p$ 
and $\eta_p$ used to describe the orientation of $\bs_p$) 
and its magnitude by the dimensionless spin parameter $\tilde a\le 1$ 
according to
\be
S_c=\tilde a\frac{GM_c^2}{c}.
\label{eq:chi_S}
\ee
We then find
\ba
&& (\Delta t)_{\rm FD}=-(\Delta t)_{\rm FD}^{\rm max}\,\tilde a\,\sin\zeta_c
\left(\frac{R_E}{b_\pm}\right)\nonumber\\
&& \times\frac{\sin\eta_c\cos\psi-\cos\eta_c\sin\psi\cos i}
{(1-\sin^2\psi\sin^2 i)^{1/2}}
\label{eq:final}
\ea
where we have introduced a fiducial unit of time
\ba
&& (\Delta t)_{\rm FD}^{\rm max}=
{R_g^2\over cR_E}=
\frac{R_g}{c}\left(\frac{R_g}{2a_\parallel}
\right)^{1/2}\nonumber\\
&& \approx 1.44\times 10^{-2} \mu{\rm s}\left(\frac{M_c}{M_\odot}\right)^{3/2}
\left(\frac{R_\odot}{a_\parallel}\right)^{1/2}.
\label{eq:t_0}
\ea
Note that $(\Delta t)_{\rm FD}$ never diverges, not even 
for $\Delta i=0$.
If one were to neglect the light deflection by the companion (as has 
been done in all previous studies, see \cite{laguna}, \cite{wex}, 
\cite{tart}, \cite{rugg}), $(\Delta t)_{\rm FD}$ would
still be given by equation (\ref{eq:delay}) but with $R$ instead of $b_\pm$.
This would lead to a divergent $(\Delta t)_{\rm FD}$ 
as $i\to\pi/2$ and $\psi\to\pi/2$.
In the limit $R\gg R_E$, eq.~(\ref{eq:final}) reduces to the
corresponding result of \cite{wex} obtained neglecting lensing.

From eq.~(\ref{eq:final}), it is clear that $(\Delta t)_{\rm FD}^{\rm max}$ 
is the {\it maximum possible} value of the frame-dragging time delay 
since $b_+\ge R_E$. This fundamental limit may be 
approached if the companion is a maximally spinning black 
hole with a certain spin orientation 
and $i=\pi/2$, $\psi=\pi/2$.
Note that for the negative image, $b_-\le R_E$ 
and $(\Delta t)_{\rm FD}$ exceeds $(\Delta t)_{\rm FD}^{\rm max}$
roughly by $R_E/b_-$, 
(neglecting various geometric factors and assuming $\tilde a=1$), 
but its magnification $A_-=(b_-/R_E)^4/[1-(b_-/R_E)^4]$ rapidly goes to zero
as the time delay increases. Thus, observations of the negative
image are not feasible in practice. 


\section{Application to Highly-Inclined Systems
\label{sect:implications}}


The effects of gravitational lensing and companion motion 
discussed in previous sections are of interest only for nearly
edge-on binaries, those with $|\Delta i|=|i-\pi/2|\ll 1$.
In particular, for the ``strong-lensing'' systems (those
with $|\Delta i|\lo R_E/a_\parallel$), the image shift at the 
orbital conjunction ($\psi=\pi/2$),
$\Delta b\sim R_E\sim a_\parallel v/c$, is of the same order as the apparent 
change of the photon impact parameter due to the companion motion,
$|\bb_0-\bR|=a_\parallel V_c/c$ [see eq.~(\ref{eq:bb_0})].
Therefore it is important to account for both lensing and companion motion
properly for such systems.


\subsection{Effect of Companion Motion
\label{subsect:cm_interp}}

For a highly inclined binary system ($|\Delta i\ll 1|$)
around the orbital conjunction ($|\Delta\psi=|\psi-\pi/2|\ll 1$),
the effect of lens motion on the various time delays studied
in previous sections has a simple interpretation, as we now show.
Expanding eq.~(\ref{eq:b0^2}) for $|\Delta i|,|\Delta\psi|\ll 1$, we find
\ba 
&& b_0 \simeq  a_\parallel\left\{\left[\Delta\psi
+(d/a_\parallel)(1+e\sin\omega)\right]^2\right.\nonumber\\
&& \left.+(\Delta i)^2\left[1+(d/a_\parallel)e\cos\omega\right]^2
\right\}^{1/2}\nonumber\\
&& \approx a_\parallel\left\{\left[\Delta\psi+(d/a_\parallel)
(1+e\sin\omega)\right]^2+
(\Delta i)^2\right\}^{1/2}.
\label{eq:b02}
\ea
Comparing with the static limit, $b_0^{\rm (stat)}=R\simeq a_\parallel
\left[(\Delta\psi)^2+(\Delta i)^2\right]^{1/2}$, we see that
$b_0|_{\psi}\simeq 
b_0^{\rm stat}|_{\psi+\Delta\psi_{\rm ret}}$,
where we have introduced a phase shift 
\ba
\Delta \psi_{\rm ret}\equiv\frac{d}{a_\parallel}(1+e\sin\omega)=
\frac{\Omega_b a_c}{c}\frac{(1+e\sin\omega)}{\sqrt{1-e^2}}.
\label{eq:psi_ret}
\ea
Clearly, in the presence of the lens motion, the photon impact parameter 
is minimal not at $\Delta\psi=0$ (exact conjunction)
but at $-\Delta \psi_{\rm ret}$, which occurs slightly
{\it before} the conjunction. This is a direct manifestation of the 
retardation effect due to the finite time it takes for the photons 
emitted by the pulsar to propagate through the binary, during which 
the lensing companion's position changes (see 
\cite{kopshaf}, \cite{kopei}).

As mentioned in \S \ref{sect:timing_no_spin}, near the conjunction 
the Shapiro delay $(\Delta t)_{\rm grav}$ is described by 
eq.~(\ref{eq:Shap_conj}). It is then obvious from (\ref{eq:geom}) 
and (\ref{eq:Shap_conj}) that lens motion affects $(\Delta t)_{\rm grav}$ 
and $(\Delta t)_{\rm geom}$ only through its effect on $b_0$ which enters
$b_\pm$ in these formulae. Thus,
\ba
&& (\Delta t)_{\rm grav}|_{\psi}\simeq 
(\Delta t)_{\rm grav}^{\rm stat}|_{\psi+\Delta\psi_{\rm ret}},\nonumber\\
&& (\Delta t)_{\rm geom}|_{\psi}\simeq 
(\Delta t)_{\rm geom}^{\rm stat}|_{\psi+\Delta\psi_{\rm ret}},
\label{eq:spin_indep_relat}
\ea
for $|\Delta i|,~|\Delta \psi|\ll 1$, where $(\Delta t)_{\rm grav}^{\rm stat}$
and $(\Delta t)_{\rm geom}^{\rm stat}$ are the Shapiro and geometric delays 
calculated in the static limit \cite{LR05}.

For the spin-dependent time delays near conjunction, 
one finds using eqs.~(\ref{eq:dtl}) and (\ref{eq:delta_t_L^lat})
\ba
&& (\Delta t)_L\simeq -\left({\Delta b_\pm\over b_0}\right)\nonumber\\
&&
{ \cos\eta_p\Delta i-
\sin\eta_p \left[\Delta\psi+(d/a_\parallel)(1+e\sin\omega)\right]\over
\Omega_p\sin\zeta_p},
\label{eq:deltatl}\\
&& (\Delta t)_L^{\rm (lat)} 
\simeq -\left({\Delta b_\pm\over b_0}\right)\nonumber\\
&&
{ \sin\eta_p\Delta i+
\cos\eta_p \left[\Delta\psi+(d/a_\parallel)(1+e\sin\omega)\right]\over
\Omega_p\sin\zeta_p\tan\chi_0}.
\label{eq:delt_L_col}
\ea
%
%
It is again clear that 
\ba
&& (\Delta t)_L|_{\psi}\simeq 
(\Delta t)_L^{\rm stat}|_{\psi+\Delta\psi_{\rm ret}},\nonumber\\
&&
(\Delta t)_L^{\rm (lat)}|_{\psi}\simeq 
(\Delta t)_L^{\rm (lat),stat}|_{\psi+\Delta\psi_{\rm ret}},
\label{eq:spin_dep_relat}
\ea
where $(\Delta t)_L^{\rm stat}$ and $(\Delta t)_L^{\rm (lat),stat}$ are the 
longitudinal and latitudinal light-bending delays calculated in the 
static approximation (\cite{RL05}).

Equations (\ref{eq:spin_indep_relat}) and (\ref{eq:spin_dep_relat}) 
show that all delays discussed here are related to their static
analogs via $(\Delta t)(\psi)\simeq (\Delta t)^{\rm stat}
(\psi+\Delta\psi_{\rm ret})$ for $|\Delta i|,~|\Delta \psi|\ll 1$. 
Thus, the effect of the companion motion 
is to simply {\it shift} the delay curves $(\Delta t)^{\rm stat}(\psi)$ 
homogeneously in phase by 
$-\Delta \psi_{\rm ret}$ without affecting the amplitude or shape of these 
curves (to $v/c$ accuracy). This phase shift is equivalent to 
moving the time delay curves calculated in the static approximation
{\it earlier} in time by 
\ba
\Delta t_{\rm ret}= \frac{\Delta \psi_{\rm ret}}{\Omega_b}
\frac{(1-e^2)^{3/2}}{(1+e\sin\omega)^2}=\frac{M_p}{M}\frac{a}{c}
\frac{1-e^2}{1+e\sin\omega}.
\label{eq:t_ret}
\ea
One can see that as advertised in \S \ref{sect:intro} this time 
shift is indeed equal to $(M_p/M)a_\parallel/c$, i.e., the light 
crossing time of the binary at conjunction weighted by the mass 
ratio $M_p/M$.
%

To illustrate our results, consider the double pulsar 
J0737-3039 system, consisting of the millisecond pulsar A with 
spin period $P_A=22.7$~ms and mass $M_p=1.337M_\odot$, and the normal 
pulsar B with mass $M_c=1.25M_\odot$ playing the role of 
the lensing companion (\cite{burg}, \cite{lyne}). 
This nearly edge-on system has $P_b=0.1023~\mbox{d}$,
$e=0.0878$, and $a=8.784\times 10^{10}~\mbox{cm}$.
The phase shift is $\Delta\psi_{\rm ret}\approx 1.2\times
10^{-3}$, and time shift $\Delta t_{\rm ret}\approx 1.4$~s.
In Fig.~1 we display the timing delays discussed in this paper
for J0737-3039 -- the combined geometric and 
Shapiro delays given by eqs.~ (\ref{eq:geom})
and (\ref{eq:Shapiro}), and the rotational lensing delays
given by eqs.~(\ref{eq:dtl}) and (\ref{eq:delta_t_L^lat}) -- 
together with the image magnification factor given by eq. (\ref{eq:mag}),
which depends on $b_0$ and is thus also affected by the lens motion. 
We choose $i=90.29^\circ$ (\cite{coles}) for this figure,
which allows lensing effects to manifest themselves (see \cite{LR05}, 
\cite{RL05}). 
To evaluate the spin-dependent delays $(\Delta t)_L$ and  
$(\Delta t)_L^{\rm (lat)}$, we adopt as an example $\zeta=50^\circ$, 
$\eta=45^\circ$ consistent with the polarization measurement of 
\cite{demorest}. Compared to the static results, all 
these curves are shifted back in time from the moment of conjunction.

\begin{figure*}
\includegraphics{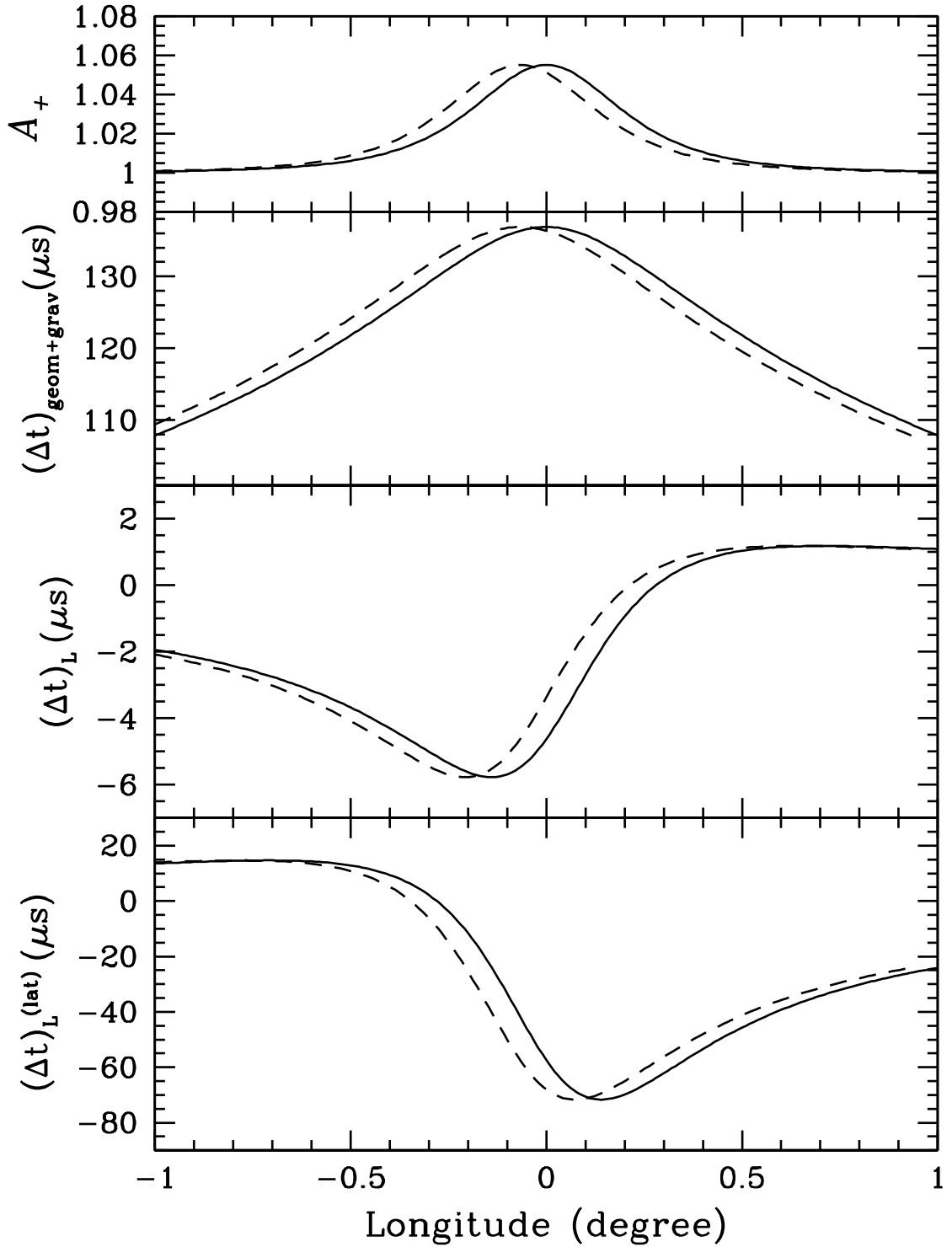}
\caption{The amplification (top panel), the combined geometric and
Shapiro delay (the second panel), the rotational-lensing delay 
(the third panel), and the latitudinal lensing
delay (the bottom panel) of the dominant (``plus'')
image of the pulsar A signal as a function of the orbital phase in the
double pulsar PSR J0737-3039 system.
The longitude is measured from the superior conjunction of pulsar A
(when A is exactly behind pulsar B). 
The inclination angle is chosen to be $90.29^\circ$.
For the third and bottom panels, the angles $\zeta=50^\circ$,
$\eta=45^\circ$ and $\tan\chi_0=0.08$ are used.
In each panel, the solid curve corresponds to the static limit
(see Fig.~1 of \cite{RL05}), while the dashed curve includes 
the effect of the companion motion.
\label{fig:pos_image}}
\end{figure*}

Although the example in Fig.~1 refers to the ``strong lensing'' 
case ($|\Delta i|\lo v/c$), we emphasize that for any highly-inclined
system ($|\Delta i|\ll 1$), the effect of the companion motion 
can be captured adequately by the ``retardation shift'' as described
above. In Table I we list $\Delta\psi_{\rm ret}$ and 
$\Delta t_{\rm ret}$ for binary pulsar systems with measured Shapiro delays.
One can see that the value of $\Delta t_{\rm ret}$ ranges from several
seconds in compact systems to several minutes in wide binaries. Detection
of the retardation effect seems easier for the latter but would 
ultimately depend on the accuracy of timing observations of a given 
system\footnote{For PSR J1909-3744 timing residuals may become as low
as $10$ ns \cite{jacoby} which should make it one of 
the best systems to probe the
retardation effect even though its $\Delta t_{\rm ret}$ is only $13$ s.}.

\begin{table*}
\caption{Binary pulsars with measured Shapiro delays. 
\label{table}}
\begin{ruledtabular}
\begin{tabular}{ l l l l l l l l l }
PSR & $P_p$\footnotemark[1], ms & $P_b$\footnotemark[2], d & $a/c$, s &
$M_c$, $M_\odot$ & $\sin i$ & $\Delta\psi_{\rm ret}$ &
$\Delta t_{\rm ret}$, s & Ref. \\
\hline
B1534+12\footnotemark[3]  & $37.9$ & $0.42$ & $7.6$ & $1.345$ & $0.975$ & $9.2\times 10^{-4}$ & $8.87$ & \cite{stairs02} \\
J0737-3039\footnotemark[3]  & $22.7$ & $0.1$ & $2.93$ & $1.25$ & $0.999987$\footnotemark[5] & $11.8\times 10^{-4}$ & $1.39$ &  \cite{lyne} \\
B1855+09\footnotemark[4] & $5.4$ & $12.3$ & $60$ & $0.23$ & $0.9995$ & $3\times 10^{-4}$ & $50.8$ & \cite{ryba} \\
J0437-4715\footnotemark[4] & $5.8$ & $5.7$ & $37.6$ & $0.24$ & $0.679$ & $4.2\times 10^{-4}$ & $32.6$ & \cite{van} \\
J1713+0747\footnotemark[4] & $4.57$ & $67.8$ & $192$ & $0.28$ & $0.9$ & $1.7\times 10^{-4}$ & $158$ &  \cite{splaver} \\
J1640+2224\footnotemark[4] & $3.16$ & $175.5$ & $369$ & $0.25$ & $0.99$ & $1.3\times 10^{-4}$ & $313$ & \cite{lomer} \\
J1909-3744\footnotemark[4] & $2.95$ & $1.53$ & $15.3$ & $0.2$ & $0.998$ & $6.4\times 10^{-4}$ & $13.4$ & \cite{jacoby} \\
J0751+1807\footnotemark[4] & $3.4$ & $0.26$ & $5.8$ & $0.19$ & $0.83$ & $14.9\times 10^{-4}$ & $5.32$ &  \cite{nice} \\
\end{tabular}
\end{ruledtabular}
\footnotetext[1]{ Pulsar spin period, $P_p=2\pi/\Omega_p$.}
\footnotetext[2]{ Orbital period of the binary, $P_b=2\pi/\Omega_b$.}
\footnotetext[3]{ Pulsar companion is a neutron star.}
\footnotetext[4]{ Pulsar companion is a white dwarf.}
\footnotetext[5]{ Value of $\sin i$ from scintillation measurements of \cite{coles}. 
Shapiro delay suggests $\sin i=0.9997$ \cite{kaspi}.}
\end{table*}


\subsection{Effect of Frame-Dragging
\label{subsect:fd_interp}}

Previous conclusions on the observability of the frame-dragging
time delay have been over-optimistic as they have always been 
obtained neglecting the light bending effect.
For example, for a binary pulsar system with $M_c=10~M_\odot$, $\tilde a=1$, 
and $a_\parallel\approx 9.4~R_\odot$ (corresponding to binary period 
of $1$ d), Wex \& Kopeikin \cite{wex} found that $(\Delta t)_{\rm FD}$ 
reaches $1~\mu$s for $|i-\pi/2|\approx 0.05^\circ$, while 
our eq.~(\ref{eq:t_0}) gives $(\Delta t)_{\rm FD}^{\rm max}\approx 0.15~\mu$s.
This difference is easily understood by noticing that for this set of 
parameters, $R_E\approx 2\times 10^4$~km is considerably 
larger than the assumed unperturbed impact parameter of the radio 
beam, $a_\parallel|i-\pi/2|\approx 5.7\times 10^3$ km. Thus, 
light bending cannot be neglected in the calculation of 
$(\Delta t)_{\rm FD}$ for this system. For binary pulsars with neutron star 
companions, $(\Delta t)_{\rm FD}$ is completely negligible 
(contrary to the claims in Refs.~\cite{tart} and \cite{rugg}).

Another problem with stellar-mass binaries is that $(\Delta t)_{\rm FD}$ 
is degenerate with the longitudinal lens-rotational delay 
$(\Delta t)_L$. This was first pointed out in \cite{wex} 
under the weak light bending approximation\footnote{Their calculation
of $(\Delta t)_L$ was in the weak lensing regime while the expression
for $(\Delta t)_{\rm FD}$ neglected lensing altogether.}, 
but it is easy to see that this degeneracy holds even in the case of 
strong lensing: $\Delta b_\pm$ satisfies the lens equation $\Delta 
b_\pm=R_E^2/b_\pm$, and the $\psi$-dependent factors in eqs.~(\ref{eq:dtl}) 
and (\ref{eq:final}) are analogous\footnote{One needs
to set $d=0$ and $b_0\to R=r(1-\sin^2\psi\sin^2 i)^{1/2}$ 
in (\ref{eq:dtl}) as eq. (\ref{eq:final}) was derived in the static limit.} 
(the only difference is in $\eta_p$ used in the
former and $\eta_c$ in the latter). As a result, these two delays
are covariant and in practice it is not possible to 
separate their contributions unless there exists some additional 
information on their relative magnitudes such as the knowledge of 
the pulsar spin-magnetic orientation from polarization measurements. 

These examples lead us to conclude that measuring frame-dragging effect
in the stellar mass pulsar binaries is 
impossible if the timing precision is to remain at the level of $1~\mu$s. 
In principle, shrinking binary
(reducing $a_\parallel$) boosts up 
$(\Delta t)_{\rm FD}^{\rm max}$, but this comes at the cost of 
reducing the binary lifetime due to gravitational wave emission \cite{laguna}.
If a pulsar is found around a high-mass black hole
(such as the one in the Galactic Center, see \cite{pfahl}) in a
highly inclined orbit, the chance of measuring frame-dragging delay
may be better: in this case $(\Delta t)_{\rm FD}^{\rm max}$ not only
is large but also strongly exceeds $(\Delta t)_L$.

\section{Discussion
\label{sect:disc}}

In this paper we have studied the combined effects of
gravitational lensing and companion motion on various
time delays in binary pulsar systems. Our study
improves upon previous work on the lensing effect
by going beyond the static approximation (\cite{LR05}, \cite{RL05}). 
We show that for highly inclined systems, the companion motion affects 
the Shapiro delay and other lensing-related delays by shifting the 
delay curves backward in time by the amount $\sim a_c/c$ (where $a_c$
is the semimajor axis of the companion).

To detect this effect 
on time delays, 
one should look for a systematic displacement of the peak of 
the Shapiro delay curve from the exact moment of the 
superior conjunction of the pulsar.
To this end, one must be able to pinpoint the exact 
moment of the conjunction and to nail down the peak of the Shapiro
delay curve.
This may not be so easy since in practice one bins the timing 
data within rather large time intervals to get a good measurement of 
the Shapiro delay. Thus, one would need repeated measurements of many
binary pulsar orbits to be able to reach the accuracy needed 
for measuring the retardation effect. 
Retardation effect on lensing delays {\it may} be detectable in 
the double pulsar system J0737-3039, but the 
feasibility of this measurement depends on the currently uncertain
inclination of the system and is compromised by the
eclipse of the millisecond pulsar near its superior conjunction 
by the magnetosphere of its companion  (\cite{lyne}, \cite{mclaugh}).

In the case of the double pulsar J0737-3039 system, the retardation 
effect due to the companion motion not only affects
timing of the millisecond pulsar but may also influence 
the interpretation of the scintillation 
measurements of this system near conjunction \cite{coles}. 
Indeed, the correlated measurements of the interstellar scintillations 
of the two pulsars provide information about the system orientation and 
properties of the interstellar turbulence only when the positions of both
pulsar projected on the sky plane are accurately known at each moment of 
time. In the static approximation the projected separation of pulsars is 
$\bR$, while it becomes $\bb_0$ [see eq.~(\ref{eq:bb_0})] when 
the retardation effect is taken into account. For the currently
inferred inclination angle from the scintillation measurement itself 
($|i-90^\circ|=0.29^\circ\pm 0.14^\circ$, \cite{coles}), 
both the retardation effect and the light bending distortion of the 
apparent path of the lensed millisecond pulsar \cite{LR05} have similar 
orders of magnitudes and contribute at the level of $\sim R_E$
to the projected pulsar separation at conjunction.
As for this system $R_E=2550$ km is not very different from the inferred 
minimum projected approach distance of the two pulsars, 
$\approx 4000\pm 2000$ km, one expects that both 
effects may at some level affect the interpretation of the correlated 
scintillation measurements of \cite{coles}. Whether this can
explain the discrepancy between the values of the system's inclination 
obtained using the Shapiro delay ($i=88.7^\circ\pm 0.9^\circ$; 
\cite{lyne}, \cite{ransom}) and scintillations is not clear 
at present. 

\begin{acknowledgements} 

RRR thankfully acknowledges the financial support by the 
Canada Research Chairs Program. DL thanks Saul Teukolsky for
discussion, and is supported in part by 
NSF grant AST 0307252 and NASA grant NAG 5-12034.

\end{acknowledgements}

\appendix

\section{Formal Derivation of the Shapiro Delay Formula for Moving Lens
\label{ap:Shapiro}}

The metric produced by the companion (mass $M_c$) in its rest frame 
(``primed'' frame) assumes the standard form (we set $G=c=1$ in this
appendix)
\be
ds^2=-(1+2\Phi)(dt')^2+(1+2\Phi)^{-1}d\bx'\cdot d\bx',
\ee
where $\Phi=-M_c/|\bx'|$, and $\bx'$ is the 3-vector measured from
$M_c$. In the ``lab'' frame (``unprimed'') comoving with the barycenter
of the binary, where the lens moves with velocity $\bv_c$,
the metric can be obtained by a Lorentz boost. To order 
${\cal O}(v_c)$, it is given by 
\be
ds^2=-(1+2\Phi)dt^2+8\Phi\,\bv_c\cdot d\bx dt
+(1-2\Phi)d\bx\cdot d\bx.
\label{eq:ds2}\ee
The above result can also be obtained using the linearized theory
of gravity (e.g., \cite{misner}): The metric coefficients
produced by a source with density $\rho(t,\bx)$ and velocity 
$\bv(t,\bx)$ are
\ba
&&g_{00}=-1+2U+{\cal O}(\epsilon^2),\\
&&g_{0i}=-4V_i+{\cal O}(\epsilon^{5/2}),\\
&&g_{ij}=\delta_{ij}(1+2U)+{\cal O}(\epsilon^2),
\ea
where $\epsilon\sim {\cal O} (v_c^2,M_c/r)$, and
\ba
&&U=\int {\rho(t-|\bx-\bx'|/c,\bx')\over |\bx-\bx'|}d^3x',\\
&&V_i=\int {\rho\bv_c(t-|\bx-\bx'|/c,\bx')\over |\bx-\bx'|}d^3x'.
\ea
Taylor expansions show that the retardation effect enters the potentials
only in the order ${\cal O} (v_c^2)$, and eq.~(\ref{eq:ds2})
is recovered (e.g., \cite{will}).

Let the position vector of the companion (lens) be $\br_c(t)$, and
that of the ray be $\brph(t)$. Along the ray, the potential
$\Phi=-M_c/|\bx'|$ should be evaluated at 
\be
\bx'=\brph'(t')-\br_c'(t')\simeq \brph(t)-\br_c(t).
\label{eq:bx'}\ee
A Lorentz transformation shows that the correction to 
eq.~(\ref{eq:bx'}) is of order ${\cal O}(v_c^2)$ and will be neglected.
The photon trajectory $\brph(t)$ satisfies $ds^2=0$, which gives
\be
dt=\left[1-2\Phi (1-2\bv_c\cdot\bk)\right] dl,
\label{eq:dt}\ee
where $d\brph=dl\,\bk$ (with $\bk$ the unit vector of photon 
propagation) and $dl=(d\brph\cdot d\brph)^{1/2}$. This corresponds to
an effective index of refraction $1+(2M_c/|\bx'|)(1-2\bv_c\cdot
\bk)$ (e.g., \cite{schbook}). 

Consider a photon emitted from the source (the pulsar) at time 
$t_e$ (near the binary conjunction), and arriving to the observer at 
$t_o$. The impact parameter of the photon is $b=b_0+\Delta b$
[see eq.~(\ref{eq:eps})]. Integration of eq.~(\ref{eq:dt}) along the 
ray path yields 
\be
t_o-t_e=|\brph(t_o)-\brph(t_e)|+(\Delta t)_{\rm geom}
+(\Delta t)_{\rm grav},
\ee
where the 1st term gives the usual Roemer delay, the 2nd term
is the geometric delay given by equation (\ref{eq:geom}), and 
the 3rd term is the Shapiro delay represented by 
\ba
(\Delta t)_{\rm grav}=\int\! dl\,
\frac{2M_c}{|\bx'|} (1-2\bv_c\cdot\bk), 
\label{eq:shap}
\ea 
where the integration is along the ray path and
$\bx'=\brph'(t')-\br_c'(t')\simeq \brph(t)-\br_c(t)$. 

Without loss of generality, let us assume 
that the photon passes through the lens plane at $t=t_l=0$.
Then the photon position vector is $\brph(t)=
\brph(0)+ct\bk$, with $\bk=\bn_0$ for $t>0$ and 
$\bk=\bn=\bn_0+\Delta\bn$, where $\bn_0$ is the unit vector 
pointing toward the observer and $\Delta\bn=
\Delta\bb/a_\parallel$.
The lens position vector is $\br_c=\br_c(0)+\bv_c t
=\br_c(0)+\bV_c t+v_\parallel t\bn_0$, 
where $v_\parallel=\bn_0\cdot\bv_c$, and ${\bf V}_c$ is the projection 
of $\bv_c$ on the sky plane. Thus, along the ray
$\bx'=\left(\bb-\bV_c t+\sigma c\Delta\bn t\right)+(c-v_\parallel)t\bn_0$, 
with $\sigma=1$ for $t<0$ and $\sigma=0$ for $t>0$, 
and $\bb=\bb_\pm=\brph(0)-\br_c(0)$. Taking the integral in (\ref{eq:shap}) 
we have to order $v_c^2$
\ba 
&& \Delta t_1=2M_c(1+v_\parallel)\ln \left(s+\sqrt{s^2+b^2}\right)\Bigl
|_{s=-\bb\cdot\bV_c}^{s=s_o(1-v_\parallel)-\bb\cdot\bV_c}
\nonumber\\ 
&&
+2M_c(1+v_\parallel)\nonumber\\ 
&& \times\ln \left(s+\sqrt{s^2+b^2}\right)\Bigl
|_{s=s_e(1-v_\parallel)-\bb\cdot (\bV_c-c\Delta\bn)}
^{s=-\bb\cdot (\bV_c-c\Delta\bn)}, 
\ea
where $s_e=ct_e,~s_o=ct_o$. Since $s_o\rightarrow \infty$, the upper
limit only amounts to a constant which can be dropped.  Thus 
to leading order
\ba
&& \Delta t_1=-2M_c(1+v_\parallel)\nonumber\\
&& \times\ln
\left(s+\sqrt{s^2+b^2}\right)_{s=s_e(1- v_\parallel)-\bb\cdot
(\bV_c-c\Delta\bn)}.
\ea 
Note that $s_e=ct_e$ is determined by
$\bn_0\cdot\left[ \br_p(t_e)+(0-t_e)c\bn\right]=\bn_0\cdot\br_c(0)
=\bn_0\cdot\left[\br_c(t_e)-\bv_c t_e\right]$, 
which gives 
$s_e=ct_e=-(1+v_\parallel)\, r_\parallel$, where
$r_\parallel=-\bn_0\cdot\br(t_e)$ and $\br=\br_p-\br_c$.  
Thus the Shapiro delay 
$(\Delta t)_{\rm grav}=(1-2v_\parallel)\Delta t_1$ is given by 
\ba
&& (\Delta t)_{\rm grav}
= -2M_c(1-\bn_0\cdot\bv_c)\nonumber\\
&& \times \ln
\left[\sqrt{(r_\parallel+\bb\cdot\bV_c-c\bb\cdot\Delta\bn)^2+b^2}
\right.\nonumber\\
&& \quad \left. -r_\parallel-\bb\cdot\bV_c+\bb\cdot\Delta\bn\right],
\label{eq:Shap_new}
\ea
where all quantities are evaluated at $t=t_e$.

There are some differences between eq.~(\ref{eq:Shap_new}) 
and eq.~(\ref{eq:Shapiro}) (apart from the trivial additive 
constant term in the latter). These differences appear in the 
expression inside the logarithm, and are of order $(v_c/c)^2$. 
This is expected since our calculation includes
only the leading-order $v_c/c$ correction to the ``static''
Shapiro delay formula, while higher-order terms are neglected.
Indeed, it is easy to show that eq.~(\ref{eq:Shap_new}) 
reduces to eqs.~(\ref{eq:Wex}) or (\ref{eq:Shap_conj}) in the 
appropriate limits. 
We conclude that both eqs.~(\ref{eq:Shapiro}) and (\ref{eq:Shap_new}) 
give the same $v_c/c$ corrections to the Shapiro delay, 
and they should in practice give the same quantitative result 
for $(\Delta t)_{\rm grav}$.


\bibliography{ms}

\end{document}